\documentclass{article}

\usepackage{times}
\usepackage{graphicx} 
\usepackage{subfigure}
\usepackage{natbib}
\usepackage{hyperref}
\usepackage{url}

\usepackage[accepted]{grcon}

\title{gr-PHYSEC: Real-time Channel-based Key Generation for Physical Layer
Secure Wireless Communications}

\begin{document}

\twocolumn[
\grcontitle{gr-PHYSEC: Real-time Channel-based Key Generation for Physical Layer Secure Wireless Communications}

\grconauthor{Jose A. Sanchez Viloria}{josesanchez2019@fau.edu}
\grconauthor{George Sklivanitis}{gsklivanitis@fau.edu}
\grconauthor{Dimitris Pados}{dpados@fau.edu}
\grconaddress{Center for Connected Autonomy and AI (CA-AI.fau.edu), Florida Atlantic University, Boca Raton, FL, USA \\ 
}
\grconkeywords{Physical layer security, channel fingerprinting,
key generation, GNU Radio, SDR, ONNX}
]


\begin{abstract}
Securing wireless communication against eavesdropping is critical, particularly in dynamic and decentralized environments. We present \textit{gr-PHYSEC}, a new GNU Radio out-of-tree (OOT) module for real-time physical-layer key generation. Unlike traditional key generation that relies on pre-shared secrets or computational complexity, our approach derives symmetric keys from the wireless channel’s inherent randomness. We embed a trained neural network within GNU Radio to extract channel features between trusted parties (Alice and Bob) during probe exchanges. These features are quantized into binary keys, reconciled via Reed-Solomon encoding, and further secured with SHA-512 hashing. The generated keys are then directly used to encrypt data. 
Real-world experiments at the FAU CAAI connected robotics testbed using ADALM Pluto software-defined radios and NVIDIA Jetson Orin  validate the approach with ground robotic platforms. Results demonstrate low key disagreement rates and strong randomness, as verified by the NIST test suite for random and pseudorandom number generators for cryptographic applications. 
This integration showcases how GNU Radio can support real-time AI-driven security solutions, pushing the boundaries of software-defined secure communication.
The source code for this project is available at:
\begin{center}
\url{https://github.com/C2A2-at-Florida-Atlantic-University/gr-PHYSEC}.
\end{center}
\end{abstract}

\begin{figure}[t]
\centering
\includegraphics[width=\columnwidth]{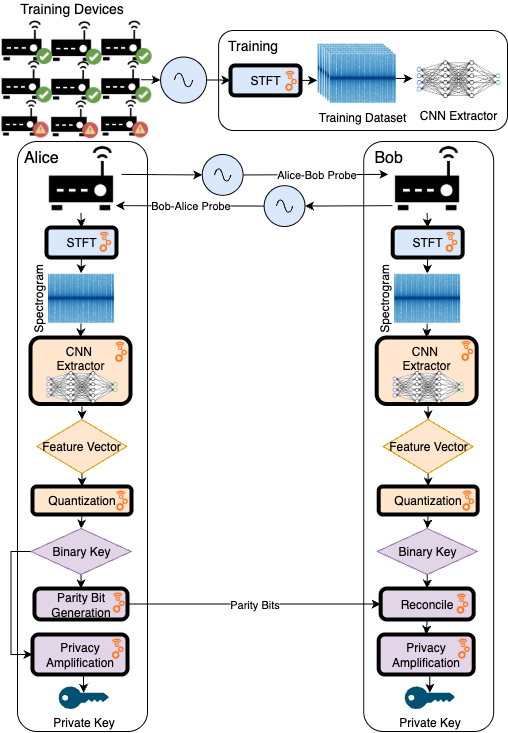}
\caption{System overview of the AI-assisted physical layer key generation system implemented in GNU Radio.}
\label{fig:System}
\end{figure}

\section{Introduction}
Wireless links expose traffic to passive and active adversaries; yet, symmetric key generation and exchange remains a difficult challenge for distributed, resource-constrained devices in infrastructure-free ad-hoc networks. Physical-layer key generation leverages channel reciprocity, temporal variation, and spatial decorrelation to derive similar features from over-the-air probes without prior information exchange \cite{zhang_access16, zhang_review16}. Prior systems have quantized received signal strength (RSS), channel state information (CSI), or channel impulse response (CIR) and reconciled errors via interactive protocols or forward error correction (FEC) \cite{liu_infocom13, zeng_infocom10, mathur_mobicom08}. In parallel, deep learning at the physical layer (PHY) has proven effective for robust feature extraction, automatic modulation classification and radio-frequency (RF) fingerprinting application \cite{oshea_tccn17, wang_china17, shen_jsac21}. 

Existing implementations of real-time key generation follow the same pipeline for channel probing, quantization, reconciliation, and privacy amplification but differ on channel sounding as well as processing techniques. A GNU radio LoRa SDR implementation demonstrated end-to-end probing, quantization, error-correction on long range based networks \cite{YingyingHu}. Other GNU Radio implementations that aimed to facilitate physical layer security techniques on GNU Radio involved a multiple-input single-output (MISO) system that uses CSI from a target receiver to inject noise as well as a single-carrier Alamouti coding system for undoing phase shifts of a pseudo-random sequence for decoding a signal \cite{RylandKevin}.

In this paper, we demonstrate for the first time real-time physical layer key generation for on-the-fly data encryption in peer-to-peer ad-hoc wireless communications. We implement and test on GPU-enabled software-defined radios GNU Radio out-of-tree (OOT) signal processing blocks for key generation between two trusted communicating parties, Alice and Bob. \ref{fig:System} depicts the end-to-end AI key generation process from training to deployment and testing. We trained a convolutional neural network (CNN) with quadruplet loss to map IQ-level spectrograms of channel probes to compact channel embeddings. Alice and Bob use their respective channel embeddings to generate a bit vector long enough to render its estimation by a brute-force attack computationally impractical no matter the computing resources of an eavesdropper. Hardware impairments, slow exchange of channel probes and mobility may impact channel reciprocity, therefore there may be differences between the generated keys. Alice and Bob follow a reconciliation algorithm based on a Reed-Solomon (RS) error correction coding technique. To minimize the amount of information leaked to Eve during the
reconciliation process, the SHA-3 is used to map the reconciled bit vector to a random bit sequence of 512 bits i.e., the desired cryptokey. 


We contribute to the software-defined radio community an open-source GNU Radio OOT project titled {gr-PHYSEC}, that implements the key generation pipeline (described in Fig. \ref{fig:System}) in real time with COTS SDR hardware, controlled by Jetson Orin GPU platforms on ground robotic platforms. We expose simple PMT-based control messages for control flow implementation, and ready-to-run flowgraphs. gr-PHYSEC includes GNU Radio OOT blocks for orchestrating channel probing between two SDRs, short-time Fourier Transform (STFT) spectrogram generation, deep learning inference from the Open Neural Network Exchange (ONNX), channel feature quantization, RS encoding/decoding for key reconciliation, and privacy amplification via SHA-512 hashing. We also developed message-oriented state machines that coordinate probing between a pair of SDRs and parity bit exchanges for the reconciliation algorithm including retransmissions and an acknowledgment protocol. The project source code is available open-source via github including the pre-trained AI feature extractor model, and Docker containers to reproduce embedded deployment on NVIDIA Jetson platforms and ADALM-Pluto SDRs.


\section{Secret Key Generation Protocol}
Figure~\ref{fig:System} depicts our proposed secret key generation protocol implemented in GNU Radio. Two legitimate/trusted parties (Alice/Bob) implement bi-directional channel probing. Each receiver generates spectrograms based on received IQ samples. Each party extracts channel embeddings based on a pre-trained CNN feature extractor that is installed at each SDR node. The neural network output is quantized to a 512-bit initial secret message. Alice uses an RS encoder, and then transmits her parity bits to Bob to reconcile. Bob combines the received parity bits with his initial secret message to reconcile bit disagreements using an RS decoder. The final step of the protocol includes privacy amplification to enhance the randomness of the desired cryptographic key  key \cite{10213440, 9598102, 9598159, zhang_review16, nist_sp80022, reed_solomon_book}.

\subsection{Channel-based feature extraction}
During bi-directional channel probing, each SDR node buffers $L{=}8192$ IQ samples. We compute an STFT using a Hamming window $N{=}256$ and 50\% overlap ($R{=}128$), forming an $M{\times}N$ log-power spectrogram ($M{=}31$) \cite{allen_stft77}. A 15-layer ResNet CNN model maps the $31{\times}256$ spectrogram input to a $512{\times}1$ element feature vector embedding using L2-norm and bounded activation for stable quantization \cite{he_cvpr16}. We use the quadruplet loss function to train the neural network. The objective is to minimize Alice$\leftrightarrow$Bob channel embeddings while maximizing the distance between Alice$\rightarrow$Eve and Bob$\rightarrow$Eve embeddings where Eve is a third SDR node playing the role of a passive eavesdropper. Our goal is to preserve channel reciprocity and boost spatial decorrelation.


\subsection{Key generation}
After channel probing and feature extraction, Alice and Bob proceed with quantizing the channel embeddings to generate an initial secret message. Due to non-perfect channel reciprocity and hardware impairments, we except that there will be bit disagreements between the two messages. Let us denote by RS(C,K) the RS encoder that takes as input
$K = L/P$ symbols (or L bits where P represents the number
of bits per symbol) and maps it into a codeword of $C = 2^P-1$ symbols. The first $K$ symbols of the RS codeword are identical to the input symbols while the remaining $S = C-K$ symbols correspond to the parity symbols. We use $C=255$ and $K=128$ and consider that Bob receives Alice's generated parity symbols with no errors. By adding $S$ parity symbols to his data, Bob's RS(C,K) decoder can detect any combination of up to and including S erroneous symbols, or locate and correct up to and including $T =\lfloor S/2\rfloor$ erroneous symbols at unknown locations. Finally, Alice and Bob apply the SHA-3-512 has function to the output bit vector from the reconciliation stage to produce an output bit vector of fixed
length, i.e., the desired cryptographic key of 512 bits. The properties of the SHA-3 hash function guarantee that even if Eve’s reconciled vector has 1-bit difference from the Alice's message, her output will be completely different than that of Alice and Bob \cite{nist_sp80022}.

\begin{figure} [t]
\centering
\includegraphics[width=\columnwidth]{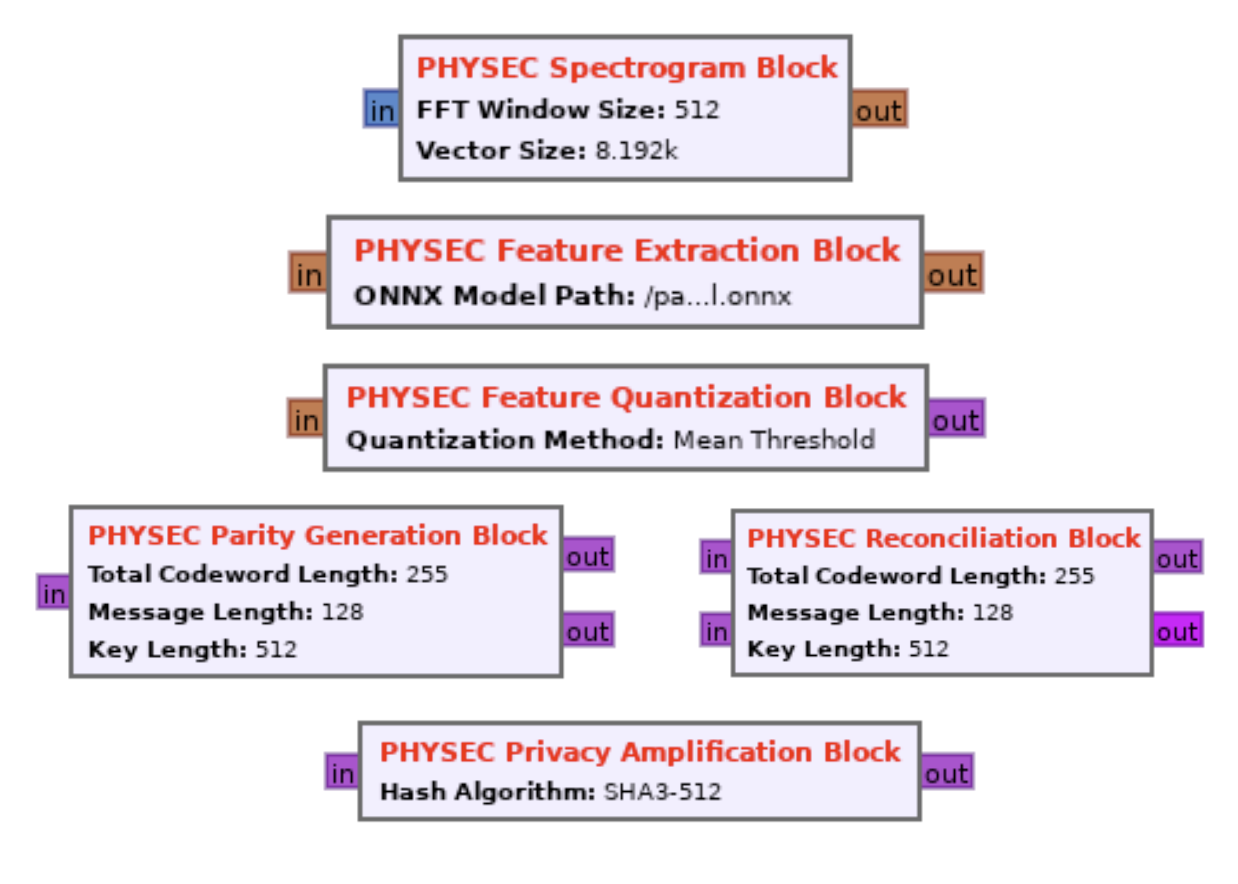}
\caption{gr-PHYSEC blocks: spectogram generation, feature extractor, feature quantization, parity bit generation, reconciliation, and privacy amplification.}
\label{fig:Blocks}
\end{figure}

\section{gr-PHYSEC}
\begin{figure*}[t]
\centering
\includegraphics[width=\columnwidth*2]{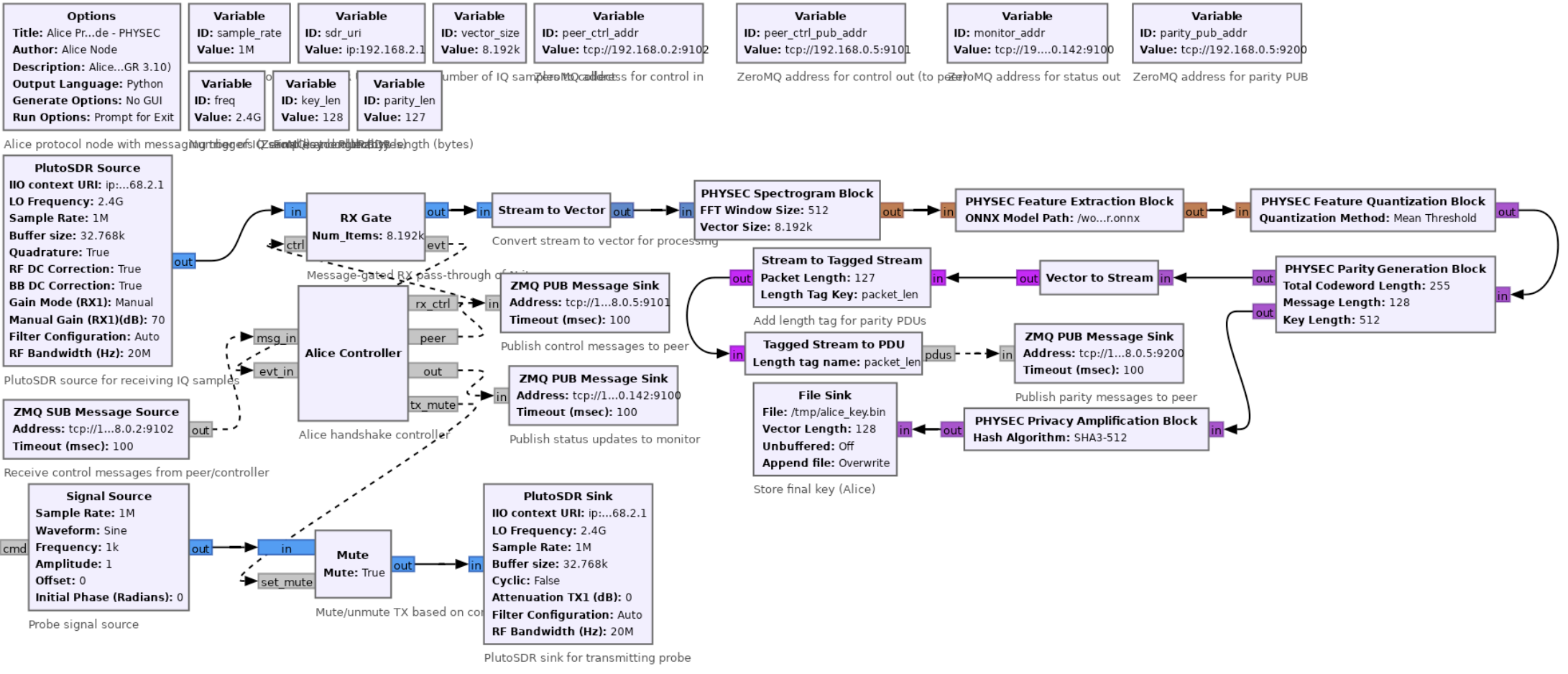}
\caption{Alice's GNU Radio flowgraph.}
\label{fig:Alice}
\end{figure*}

\begin{figure*}[t]
\centering
\includegraphics[width=\columnwidth*2]{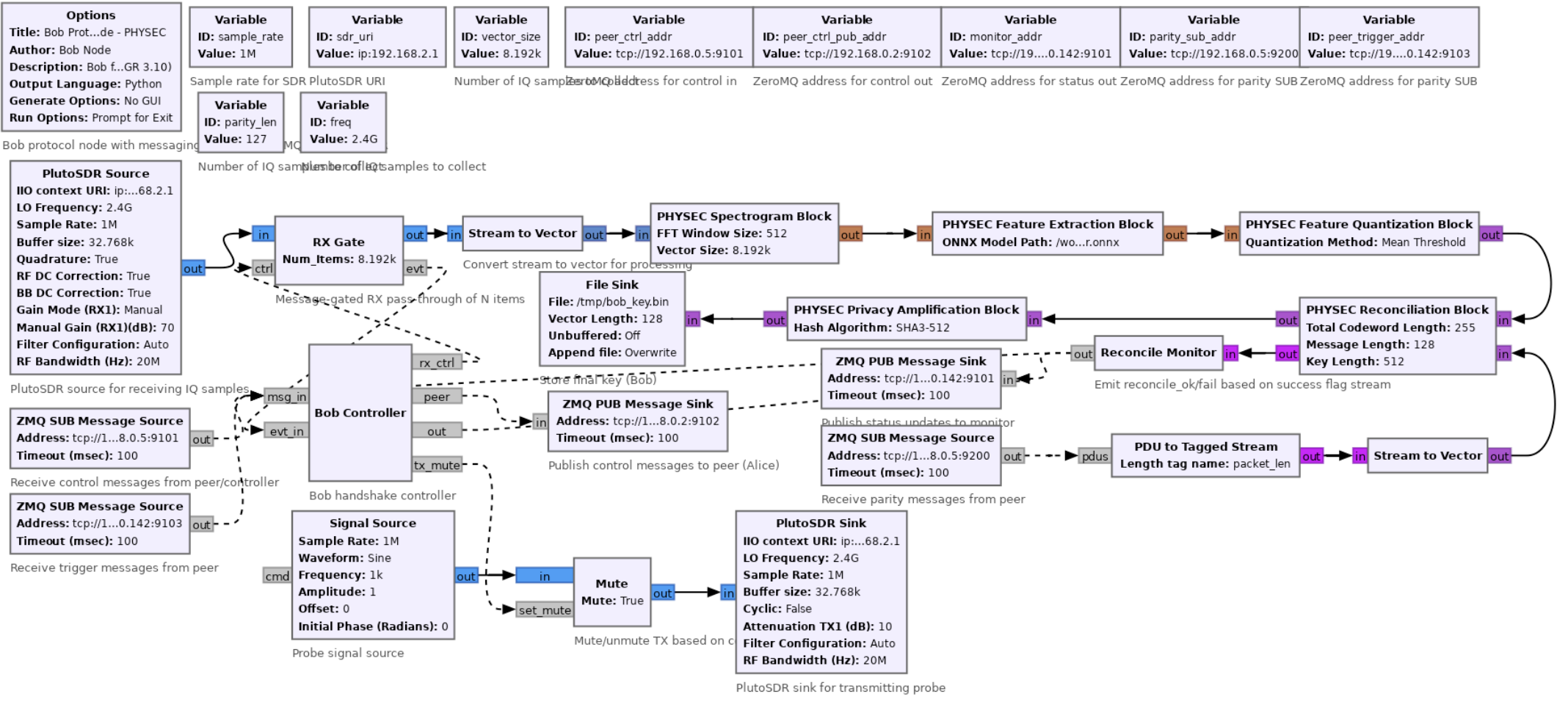}
\caption{Bob's GNU Radio flowgraph.}
\label{fig:Bob}
\end{figure*}

We developed six GNU Radio OOT blocks to enable secret key generation as well as additional blocks to control channel probing and information exchange between the two SDRs. Figure \ref{fig:Blocks} shows the individual blocks while Figures \ref{fig:Alice} and \ref{fig:Bob} show the GNU Radio flowgraphs for Alice and Bob, respectively.

\noindent\textbf{Spectrogram Generation.} A Python block pre-processes the collected IQ samples using STFT (NumPy/FFT) and log-power scaling to produce spectrograms.

\smallskip
\noindent\textbf{Feature Extraction.} This block loads the ONNX-formatted ResNet CNN model for channel-based feature extraction and runs ONNX runtime inference to produce channel embeddings. On Jetson Orin, the CUDA execution provider accelerates inference.

\smallskip
\noindent\textbf{Feature Quantization.} This block uses the arithmetic mean of each channel embedding as a threshold to quantize channel features between 0 and 1 and outputs the initial secret message at Alice and Bob.

\smallskip
\noindent\textbf{Reconciliation.} Two GNU Radio blocks (Parity Generation and Reconciliation) implementing the reconciliation algorithm. We used a Python/C++ wrapper for a Reed-Solomon library. Alice generates parity bits using her quantized channel feature vector and an RS(255,128) encoder. Bob uses Alice's parity bits and his locally generated quantized channel feature vector with an RS(255,128) decoder and reconciles to Alice.

\smallskip
\noindent\textbf{Privacy Amplification.} This block implements SHA hashing of the reconciled secret messages. We allow the user to toggle between different variations of the SHA-3 algorithm and support different output key sizes such as 128, 256, and 512 bits.

\smallskip
\noindent\textbf{Controller.} A lightweight state machine coordinates: (1) Bob$\to$Alice probe request, (2) Alice TX probe \& Bob Rx, (3) Alice$\leftarrow$Bob probe request, (4) Bob TX probe \& Alice Rx, (5) Alice  TX parity \& Bob Rx, (6) Reconciliation success/failure acknowledgment. All transitions use PMT, PlutoSDR source/sink blocks for probe transmission and ZMQ Pub/Sub for information exchange. We developed an Rx gate GNU Radio block to control the flow of received samples through the PlutoSDR Source at Bob or Alice only when Alice or Bob is transmitting. 

\begin{figure} [t]
\centering
\includegraphics[width=\columnwidth]{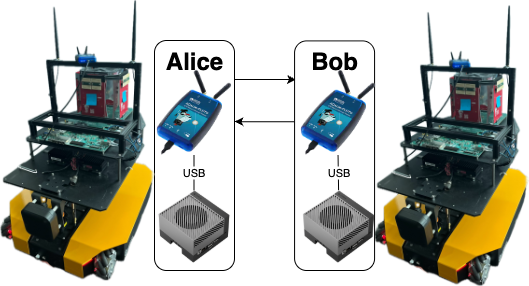}
\caption{Alice and Bob experimental setup.}
\label{fig:AliceBobRobots}
\end{figure}

\section{Experimental Setup}

We collected IQ samples from pairs of SDR devices and a third SDR playing the role of a passive eavesdropper in an indoor lab setup at FAU CAAI. 
We collected probe exchanges between Alice and Bob while Eve listened to each of their exchanges.
We trained the ResNet CNN model on an NVIDIA A100 GPU for channel-based feature extraction. We exported the CNN model to ONNX format and deployed and tested it on both Jetson Orin and Jetson Nano devices. Our experiments consider two trusted parties (Alice/Bob) at approximately $\sim$1\,m apart and then moving one of the nodes at no more than a $\sim$5\,m range. Each node is comprised of a Jetson Orin interfaced to an ADALM-Pluto SDR mounted on a robotic agent as depicted in Fig. \ref{fig:AliceBobRobots}. We built a Docker container on each of the robotic agents which contains all requirements for running the gr-PHYSEC blocks. The PlutoSDR at each agent was tuned to a carrier frequency of 2.485\,GHz, and sample rate of $1$\,MSps. We also developed a key performance indicator (KPI) monitor app which triggers the nodes to initiate execution of the secret key generation protocol and collects and visualizes information from each stage such as IQ samples and spectrograms from channel probing, quantized channel feature vectors, and reconciliation success. We use the collected information to calculate bit disagreement rate (BDR), key generation success rate, and key generation time.

\section{Results}\label{sec:results}
\subsection{Bit disagreement and reconciliation}
\begin{figure} [t]
\centering
\includegraphics[width=\columnwidth]{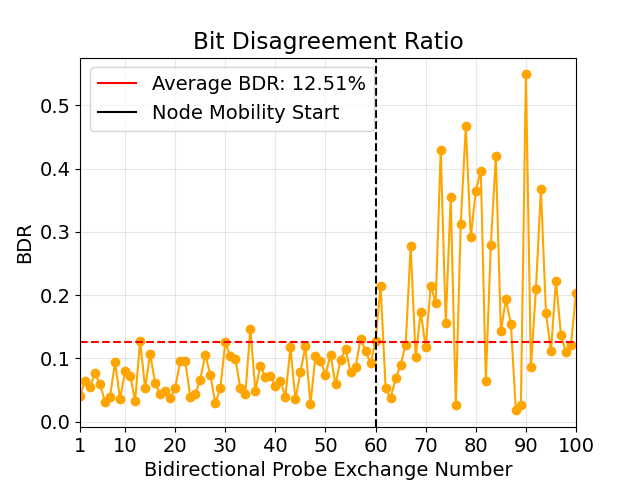}
\caption{Key generation success rate.}
\label{fig:BDR}
\end{figure}

\begin{figure} [t]
\centering
\includegraphics[width=\columnwidth]{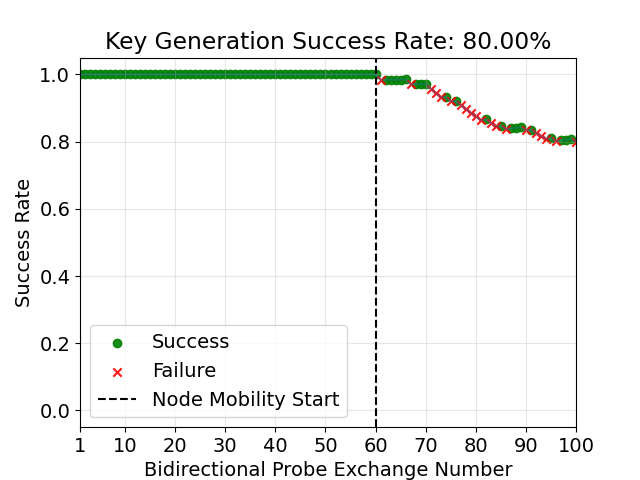}
\caption{Bit disagreement ratios at every exchange.}
\label{fig:SuccessRate}
\end{figure}

Across 100 real-time key generations, BDR between Alice/Bob keys typically lies in the 2--15\% range. Changes in orientation and motion between the two robotic agents increase BDR to $\sim$ 56\% (the representative median is $\sim$ 12. 51\%) well within the RS correction capabilities as seen in Fig. \ref{fig:BDR}. With RS$(255,128)$, reconciliation is successful in most attempts, resulting in a key generation success rate of 80\% as depicted in Fig. \ref{fig:SuccessRate}. The effects of changes in orientation and motion after epoch 60 are shown in both Fig. \ref{fig:BDR} and \ref{fig:SuccessRate}. Shorter RS codes (e.g., RS(191,128)) reduce parity exchange overhead but also success rate, as demonstrated in our prior physical layer security studies \cite{10213440, 9598102, 9598159}.

\subsection{Key randomness (NIST STS)}
We evaluate the randomness of the generated keys using the NIST Statistical Test Suite \cite{nist_sp80022} to verify whether the SHA-3 output is correlation-free and does not contain any exploitable patterns among the constituent bits. The NIST software is composed of fifteen distinct tests. In this work, we report on a representative subset commonly used in key generation studies: Monobit (Frequency), Frequency Within a Block, Runs, Longest Run of Ones in a Block, Discrete Fourier Transform (Spectral), Non\textendash Overlapping Template Matching, Serial, Approximate Entropy, and Cumulative Sums. 

Each test treats a reconciled, privacy\textendash amplified cryptokey as one binary sequence. We average $p$-values across tests and calculate key pass rates over 100 independent generated keys. Results are summarized in Table~\ref{table:NIST_1}.

\begin{table}[H]
\begin{center}
\caption{NIST statistical test suite scores.}
\begin{tabular}{ |p{2.8cm}|p{2.2cm}|p{2.2cm}|  }
 \hline
 \multicolumn{3}{|c|}{Average test scores and generated keys} \\
 \hline
Test Name & Average \newline Test Score & Passing \newline Percentage \\
 \hline
Monobit   & 0.51 & 99\% \\
 \hline
Frequency Within Block & 0.524 & 98\%  \\
 \hline
Runs & 0.508 & 100\%\\
 \hline
Longest Run Ones in a Block & 0.211 & 84\%  \\
 \hline
Discrete Fourier Transform & 0.477 & 98\%  \\
 \hline
Non Overlapping Template Matching & 0.587 & 89\%  \\
 \hline
Serial & 0.527 & 99\%  \\
 \hline
Approximate Entropy & 0.517 & 100\%   \\
 \hline
Cumulative Sums & 0.508 & 99\% \\
 \hline
\end{tabular}
\label{table:NIST_1} 
\end{center}
\end{table}

\subsection{Latency and key generation rate}
The average key generation time is $\approx$ 1299\,ms in our embedded setup using a Jetson Orin as a compute platform. The estimated time include wait times of $\approx$400\,ms that are put in place to implement bidirectional probe exchanges. Faster/optimized control links, parallelized probe exchange, or TensorRT/quantized CNNs could reduce latency \cite{onnxruntime, oshea_tccn17}.

\begin{figure} [t]
\centering
\includegraphics[width=\columnwidth]{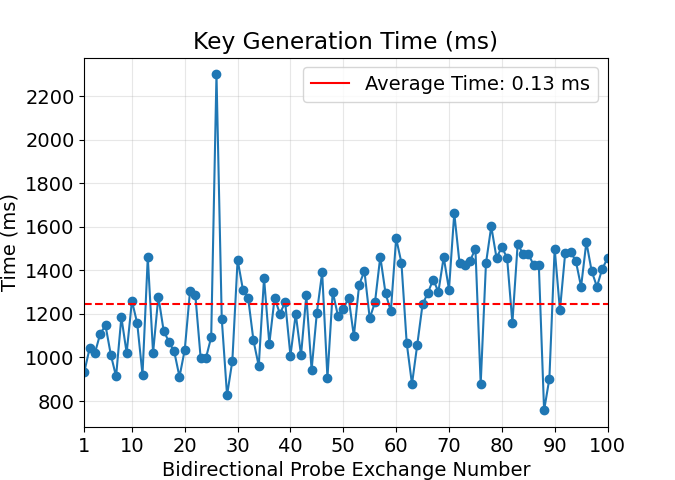}
\caption{Key generation time in milliseconds.}
\label{fig:Time}
\end{figure}

\section{Reproducibility}
The OOT blocks, example flowgraphs, and ONNX model are available at \url{https://github.com/C2A2-at-Florida-Atlantic-University/gr-PHYSEC}. We include installation notes for the gr-PHYSEC library and ONNX Runtime on Jetson platforms, example flowgraphs with control flow for running Alice and Bob SDR nodes as well as the KPI monitor which provides real-time information on probe exchanges and key performance indicators such as BDR, key generation success rate and key generation time. We developed a Docker container with the requirements to run on an ARM processor, and block-level parameter defaults. Blocks are self-contained and can be swapped (e.g., use different waveforms for channel probing).

\section{Conclusion and Future Work}
gr-PHYSEC delivers a practical, reproducible real-time physical layer key generation pipeline implemented in GNU Radio. A secret key generation protocol was implemented to generate a cryptokey between two authenticated nodes, Alice and Bob. Experimental bidirectional links were designed on low-cost COTS SDRs interfaced with edge AI inference platforms. The established links experienced a variety of channel conditions due to node mobility. Future work will focus on increasing key generation success rate and reducing key generation time through appropriate MAC-layer design as well as assessing protocol security that could be challenged by the relative position of Eve to Alice/Bob and the sophistication of Eve.

\begin{figure} [t]
\centering
\includegraphics[width=\columnwidth]{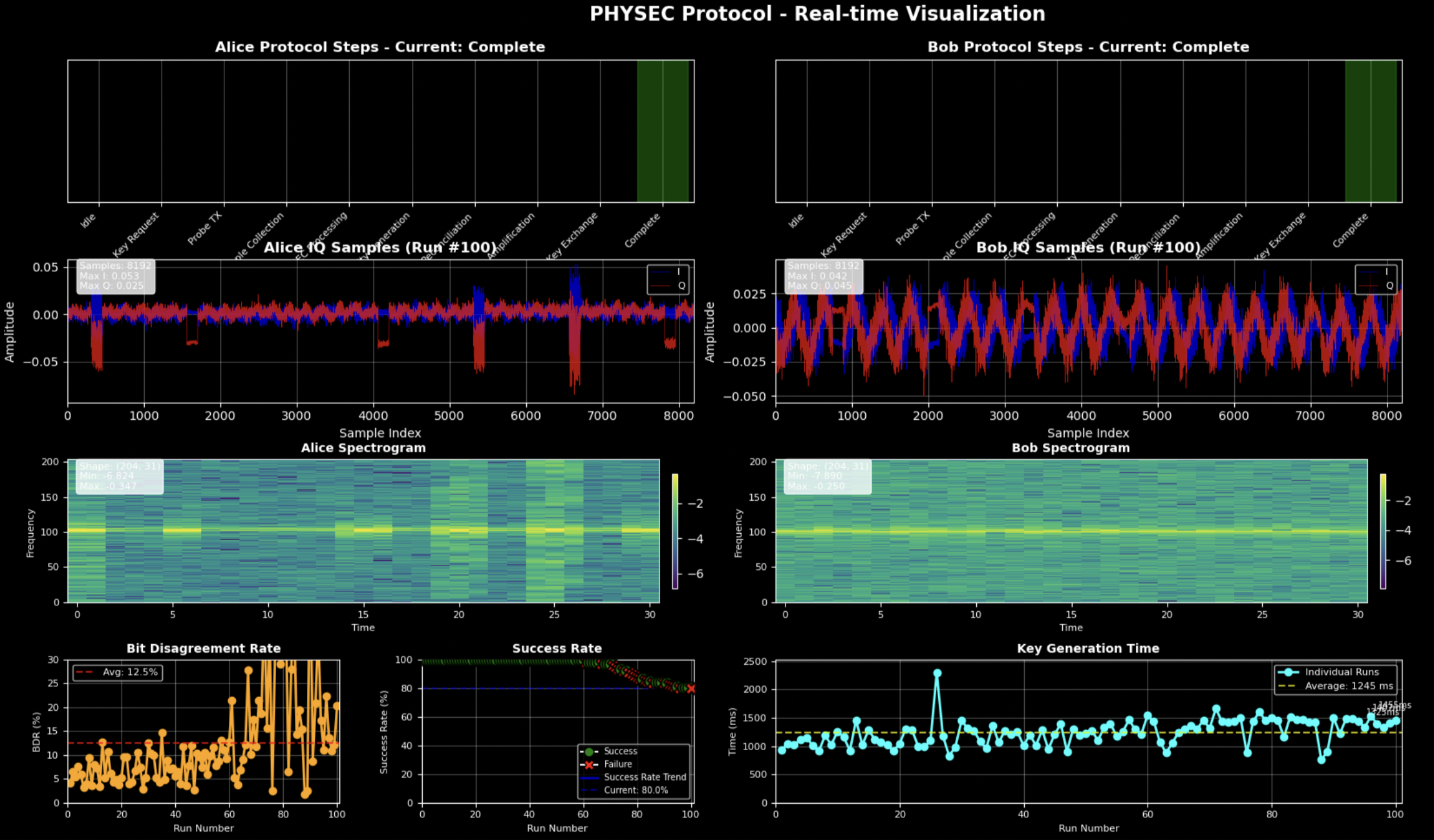}
\caption{Key Performance Indicators (KPI) monitor.}
\label{fig:Monitor}
\end{figure}

\bibliography{example_paper}
\bibliographystyle{grcon}

\end{document}